\documentclass[lettersize,journal]{IEEEtran}
\usepackage{amsmath,amssymb,amsfonts}
\usepackage{algorithmic}
\usepackage{algorithm}
\usepackage{array}
\usepackage[caption=false,font=normalsize,labelfont=sf,textfont=sf]{subfig}
\usepackage{textcomp}
\usepackage{stfloats}
\usepackage{url}
\usepackage{verbatim}
\usepackage{graphicx}
\usepackage{cite}
\usepackage{booktabs}
\usepackage{xcolor}
\usepackage{amsmath}
\usepackage{epstopdf}

\begin{document}

\title{Analytical Model of NR-V2X Mode 2 \\ with Re-Evaluation Mechanism}

\author{
\IEEEauthorblockN{
Shuo Zhu\IEEEauthorrefmark{1},
and Siyu Lin\IEEEauthorrefmark{1}
}
\IEEEauthorblockA{\\
\IEEEauthorrefmark{1}School of Electronic and Information Engineering, Beijing Jiaotong University, Beijing 100044, China \\
Siyu Lin (Corresponding Author) Email: sylin@bjtu.edu.cn
}

\thanks{This work was partly supported by the National Natural Science Foundation of China (No. 61971030) and Technology Research Programme of China Academy of Railway Sciences(Grant No. 2019YJ077).}
}

\markboth{Journal of \LaTeX\ Class Files,~Vol.~14, No.~8, August~2021}%
{Shell \MakeLowercase{\textit{et al.}}: A Sample Article Using IEEEtran.cls for IEEE Journals}

\IEEEpubid{0000--0000/00\$00.00~\copyright~2021 IEEE}

\maketitle

\begin{abstract}
Massive message transmissions, unpredictable aperiodic messages, and high-speed moving vehicles contribute to the complex wireless environment, resulting in inefficient resource collisions in Vehicle to Everything (V2X). 
In order to achieve better medium access control (MAC) layer performance, 3GPP introduced several new features in NR-V2X. One of the most important is the re-evaluation mechanism. It allows the vehicle to continuously sense resources before message transmission to avoid resource collisions. 
So far, only a few articles have studied the re-evaluation mechanism of NR-V2X, and they mainly focus on network simulator that do not consider variable traffic, which makes analysis and comparison difficult.
In this paper, an analytical model of NR-V2X Mode 2 is established, and a message generator is constructed by using discrete time Markov chain (DTMC) to simulate the traffic pattern recommended by 3GPP advanced V2X services.
Our study shows that the re-evaluation mechanism improves the reliability of NR-V2X transmission, but there are still local improvements needed to reduce latency.

\end{abstract}


\section{Introduction}
\label{section_Introduction}


Vehicle to Everything (V2X) is regarded as a key technology for the implementation of Intelligent Transportation System (ITS). It enables vehicles to communicate with adjacent vehicles and roadside units, thereby fostering an efficient and proactive safety environment. 
3GPP has published the NR-V2X standard based on 5G NR (New Radio) in Release 16 to further meet the stringent requirements of advanced V2X services, such as Vehicles Platooning, Advanced Driving, Extended Sensors, Remote Driving. 

NR-V2X introduces several new features at the Medium Access Control (MAC) layer \cite{Shin2023LTEtoNR}. One of the most distinguished from LTE-V2X (Long Term Evolution) is re-evaluation mechanism. The re-evaluation mechanism enables the vehicle to conduct multiple sensing before the message is transmitted, and once a vehicle detects that another vehicle has declared the selected resource before message transmittal, it performs resource reselection to avoid potential packet collisions. 

To the best of our present knowledge, most studies of NR-V2X mode 2 do not discuss new features of the MAC layers, only a few studies have evaluated the impact of re-evaluation mechanisms \cite{Lu2024LinkNR,Bankov2023NRtraffic}, which triggered our motivation for this study. 
At the same time, the existing methods to analyze the performance of re-evaluation mechanisms are basically based on network simulation or field experiment, such as NS-3, 5G-LENA \cite{Molina2022Reevaluation}. However, these studies failed to consider the variable traffic scenarios in V2X. Moreover, the results from network simulators are often non-replicable or non-comparable, which makes it difficult to uncover more details regarding the basic model and assumptions.
Some previous research obtained analytical expressions of transport mechanisms of LTE-V2X and NR Vehicle to Network (V2N) through model analysis\cite{Wang2022fixed, Bal2023E2E}.

In the paper \cite{WNBA2022DTMCV2X}, an analytical model of LTE-V2X semi-persistent scheduling (SPS) and IEEE 802.11p carrier sense multiple access with collision avoidance (CSMA/CA) based on DTMC is proposed. By computing the steady-state probability of the model, the average delay, channel occupancy and other indicators of system performance are obtained. By combining discrete-time Markov chains (DTMC) model with queuing theory and other models, communication in vehicle networks under unsaturated network conditions can be effectively analyzed. Therefore, we consider utilizing analytical modeling to reduce the cost of solving the model to cover as much of the network parameter space as possible.
\IEEEpubidadjcol 

The main contributions of this paper are to propose an analytical model of NR-V2X mode 2 with re-evaluation mechanism based on DTMC. This model takes into account the re-evaluation mechanism that employs the all-slot strategy and the device-level queue, along with the variable traffic scenarios recommended by 3GPP advanced V2X services. Numerical evaluation is used to provide in-depth insights and evaluations of derived performance metrics, such as latency and collision probability.
The simulation results indicate that the re-evaluation mechanism can enhance the reliability of NR-V2X transmission without substantially increasing the transmission latency. Additionally, the potential optimization directions of this mechanism are revealed.






The rest of this paper is organized as follows. We present the Model in Section II. In Section III, the steady-state probability of the Model is derived. Simulation results are given in Section IV and the paper is concluded in Section V.

\section{System Model}
\label{section_SystemModel}

\begin{figure}
	\centering
	\includegraphics[height=2.57cm,width=8.471cm]{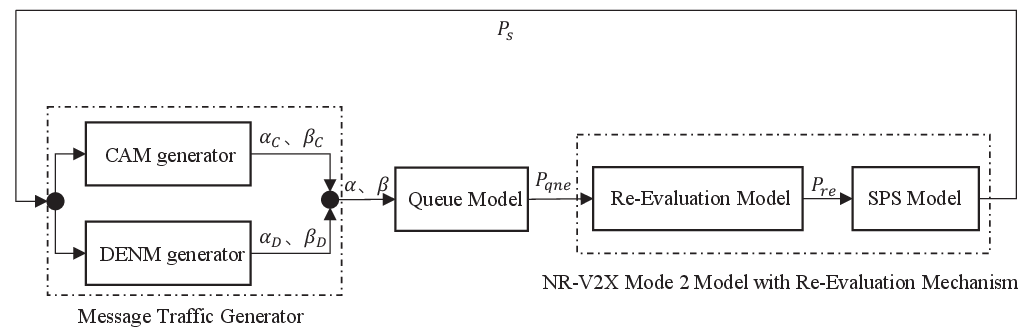}
	\caption{System Model}
	\label{system_model}
\end{figure}

This section presents all of the DTMC models for the system and illustrates the dependence among them, as shown in Fig.\ref{system_model}. Considering the real-world communication scenario, we first establish Cooperative Awareness Message (CAM) and Decentralized Environmental Notification Message (DENM) message generator to construct different traffic intensities. We refer to as Message Traffic Generator. Subsequently, we establish a Queue model to receive mixed traffic. The Re-Evaluation model and SPS model jointly realize the final dispatch of messages.

\subsection{DTMC Model for Message Traffic Generator}

\label{section_Generator}
\begin{figure}
	\centering
	\includegraphics[height=6.145cm,width=8.471cm]{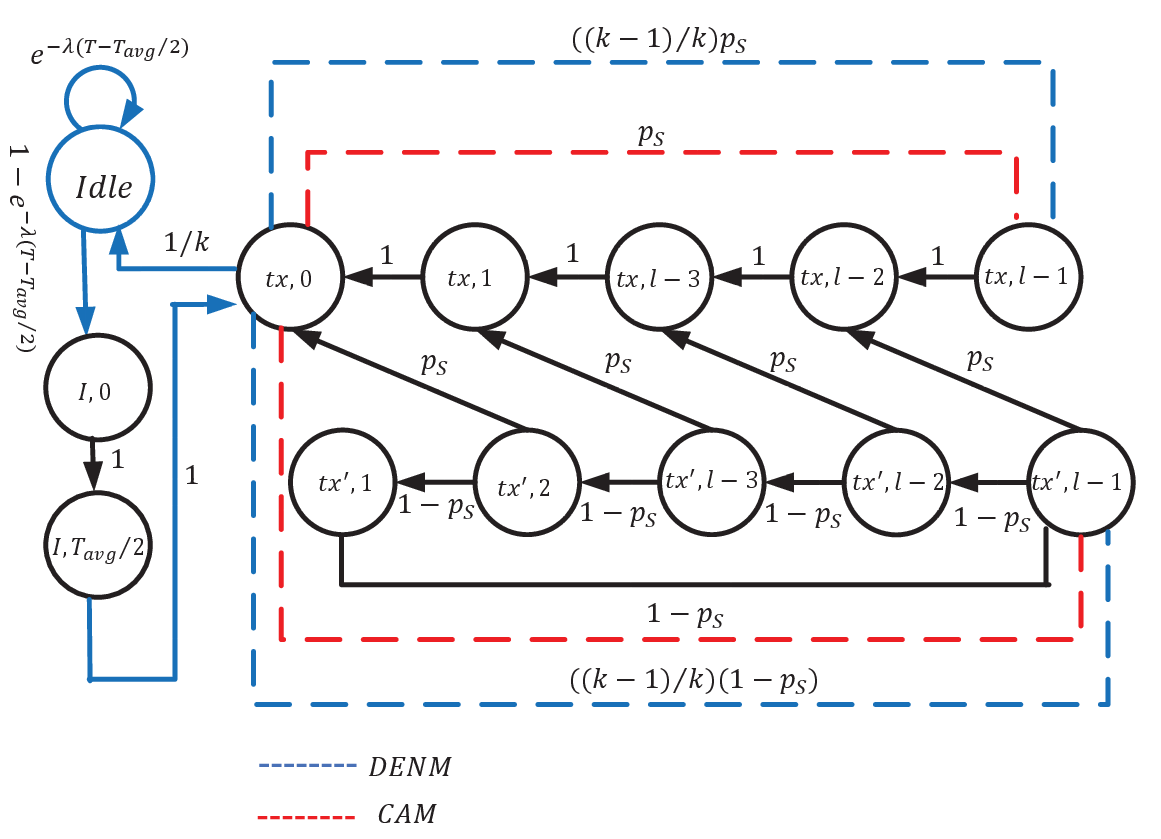}
	\caption{Message Traffic Generator}
	\label{fig_message_generator}
\end{figure}

In order to mimic the traffic flow proposed in \cite{3gpp2019service} as closely as possible, We set up CAM and DENM message generator models. 
Since they share some of the same characteristics, we show them in the same diagram, as shown in Fig. \ref{fig_message_generator}. 
The solid black line represents the state transition shared by the two traffic generators. The blue dashed line on the outer circle indicates the state transition mode exclusive to DENM, and the red dashed line on the inner circle indicates the state transition mode exclusive to CAM.

The state of both types of messages in the generator is represented by $\left ( i,j \right ) $, where $j\in \left ( 0,l-1 \right ) ,l\in \left ( T_{C},T_{D} \right )$ .If the message is transmitted successfully, the status $i$ will be $tx$, otherwise it will be $tx'$. $P_S$ is the probability of message scheduling success. 

A fixed inter-packet arrival time model is employed for periodic CAM messages, where the inter-packet arrival time $T_{C} \in [10, 30, 100] ms$ in accordance with the traffic model \cite{3gpp2019service}. $\left (tx,0 \right) $ represents the starting state in which CAM messages are successfully transmitted. Otherwise, the CAM message will wait in state ${tx}'$  until the resource is successfully scheduled.

The inter-packet arrival time of DENM traffic is aperiodic. To better support the diversity of the use case requirements, a time constant plus an exponential distribution of time variable is used to simulate the inter-arrival time. A new \emph{Idle} state is added as a trigger for the Poisson process of message generation, and the mean of the time variable is $ \left\{10, 50 \right\} ms$. Then, a one-way Markov chain is employed to represent the time constant $T_{avg}$ , where  $T_{avg} \in [20, 100]  ms$. Thus, the inter-arrival time can be expressed as $T_{D} = T_{avg} + E[X], X\backsim \mathrm{exp}(T_{avg}/2)$. To ensure the reliability of aperiodic messages, DENM messages are periodically repeated $k-1$ times at $\Delta T_{D}$ intervals after triggering generation \cite{Romeo2020kDENM}. This is consistent with the repetitive flow of CAM, and thus it is merged with CAM messages. Eventually, the generator will capture two types of DENM traffic: one is the event-triggered aperiodic messages, and one resulting from retransmissions.

\begin{figure}
	\centering
	\includegraphics[height=2.824cm,width=7.471cm]{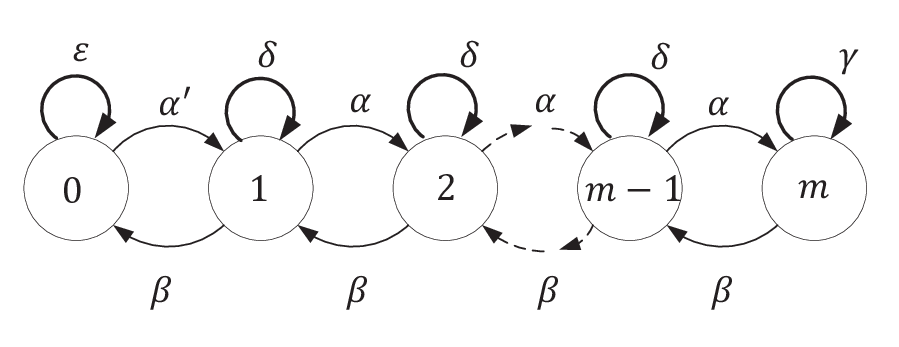}
	\caption{Queue Model}
	\label{fig_buffering_queue}
\end{figure}

Both types of messages enter the Queue Model at the device level, as illustrated in Fig.\ref{fig_buffering_queue}, upon their generation. 
Let $P_{qe}$ denote the probability of the queue being empty, $P_{qne} = 1-P_{qe}$. When the queue is empty, due to the accumulation of new messages, the transition probability of the queue from \emph{State} (0) to \emph{State} (1) is $\alpha'$. Thereafter, the transition probability of increasing the queue length is $\alpha$, and the probability of decreasing the queue length is $\beta$. 

\subsection{DTMC Model for NR-V2X Mode 2 with Re-Evaluation Mechanism}
\label{section_Mechanism}

\begin{figure}
	\centering
	\includegraphics[height=12.145cm,width=8.471cm]{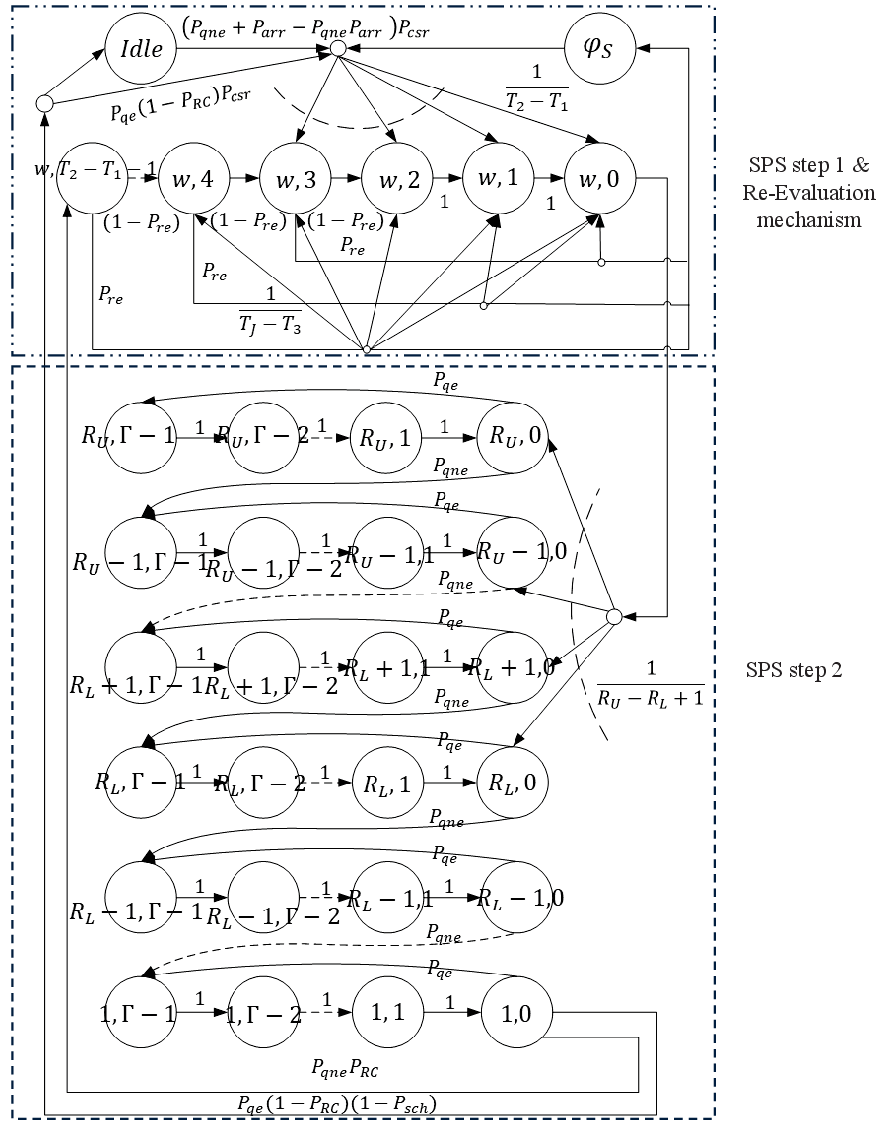}
	\caption{NR-V2X Mode 2 Model with Re-Evaluation Mechanism}
	\label{fig_reevaluation_model}
\end{figure}
NR-V2X mode 2 adopts the semi-persistent scheduling (SPS) scheme in MAC layer resource scheduling. The SPS scheme defines a 2-step algorithm for resource allocation. The vehicle excludes resources that are already occupied in \emph{step} 1, and then randomly selects an available candidate resource in \emph{step} 2. On this basis, NR-V2X mode 2 introduces the re-evaluation mechanism. The vehicle will perform the \emph{reassessment check} described in Standard\cite {3gpp2022MAC} during the \emph{step} 2 of SPS to avoid a possible collision.

Fig.\ref{fig_reevaluation_model} shows the Model of NR-V2X mode 2 with Re-Evaluation Mechanism, which is now divided into two parts, the first part describes SPS \emph{step} 1 and the re-evaluation mechanism, and the second part describes SPS \emph{step} 2.

Considering the unsaturated state of the network space, we set $idle$ as the initial state of the model. The model first performs SPS \emph{step} 1. During this step, it excludes resources that are already occupied and then selects a Resource Block (RB) within the selection window for transmission. The probability of generating a new message is $(P_{qne}+P_{arr}-P_{qne}P_{arr})P_{CSR}$. The meaning of each symbol is explained in Section \ref {section_SteadyState_Probabilities}. The model randomly selects a time index within the selection window and then enters the waiting state $(w, j)$, $j \in (0, T_2-T_1-1)$. Here, $T_1$ and $T_2$ are two parameters determined by \cite{3gpp2021data}. $T_2$ is equivalent to the packet delay budget (PDB), $T_1$ represents the processing time for resource selection. 

Then the system performs the re-evaluation mechanism during the waiting process. Assuming that the current waiting state is $(w, T_j)$ The system must complete the resource re-evaluation work before $T_j-T_3$, where $T_3$ is the necessary time for the system to perform resource awareness and selection. If a collision risk is detected in the selected RBs, SPS \emph{step} 1 is re-executed and transfers to the new waiting state. Therefore, the transition probability of the system waiting state is 1 for $(w,0)$ and $(w,1)$ and $(1-P_{re})$ for $(w,j), j \in (2,T_2-T_1-1)$, where $P_{re}$ is the probability that the system will reselect after the re-evaluation.

The second part of the model simulates the periodic transmission of messages. Resource Reservation Interval (RRI) defines the periodic interval of messages, and Reselection Counter (RC) represents the number of periodic transmissions. After each transmission, RC will be reduced by 1, and the system will reselect resources after RC is reduced to 1.The range of RC is $[R_L, R_U]$, so the probability of RC random selection is $1/(R_U-R_L+1)$.

The system enters the resource scheduling state $(i, j), i\in(1, R_U), j\in(0, T_2-1)$. For each state $(i, 0)$, the system has the opportunity to schedule resources, $i$ representing the current RC value, and the system uses this state to send control information related to semi-persistent scheduling. Therefore, the probability of successful resource scheduling is $P_S=\sum_{i=1}^{R_u}\varphi_{i,0}$. If the queue is not empty at this time, the packet data is transmitted at the same time, and then decremented $i$, and the system waits for an RRI to continue the next transmission. If the queue is empty, the system similarly waits for an RRI, but maintains its RC value until the next packet transmission. When the resource scheduling status changes to $(1, 0)$, the system may retain the resource with probability $P_{RC}$, or re-execute the resource reservation procedure to select a new resource. The probability of reselection is $1-P_{RC}$. The probability that the system retains the original resource is $P_{qne}P_{RC}$, and the probability that the system re-enters the scheduling mechanism is $P_{qne}(1-P_{RC})P_{CSR}$.

\section{Steady-State Probabilities}
\label{section_SteadyState_Probabilities}
The steady-state probability is the key to support the subsequent performance analysis. In this section, the steady-state probability of the model with $\mathit{N}$ vehicles is presented. Firstly, the message traffic generator and device level buffer queue are solved, and the steady-state probability of the NR-V2X re-evaluation mechanism model is obtained according to the solution results.

\subsection{Mode for Message Traffic Generator}
Let $\varphi_{i,j}^{CAM}$, $\varphi_{i,j}^{DENM}$ represent the steady-state probability results of CAM and DENM message generators in different states, where $i \in \left(tx, tx'\right), j \in (0, l-1), l \in (T_C, T_D)$.

For CAM generator, when $i=tx$, each state, except for state $\varphi_{tx, T_C-1}^{CAM}$, consists of two states, $tx$ successfully transmitted and $tx'$ then successfully transmitted:
\begin{equation}
\begin{aligned}
\label{e_cam_tx_send}
\varphi_{tx, j}^{CAM} & = P_S\left(\varphi_{tx, 0}^{CAM}+\sum_{z=j+1}^{T_C-1} \varphi_{tx', z}^{CAM}\right),
\\ & \quad for \ j \in\left[1, T_C-2\right],
\end{aligned}
\end{equation}
When $i=tx'$, the state is given by:
\begin{equation}
\begin{aligned}
\label{e_cam_tx'_send}
\varphi_{tx', j}^{CAM}
& =\varphi_{tx, 0}^{CAM} \frac{\left(1-P_S\right)^{T_C-j}}{\left[1-\left(1-P_S\right)^{T_C-1}\right]}, 
\\ & \quad for \ j \in\left[0, T_C-1\right].
\end{aligned}
\end{equation}

By the property that the sum of the steady-state probabilities of DTMC is one, $\varphi_{tx, 0}^{CAM}$ can be obtained as:
\begin{equation}
\begin{aligned}
\label{e_cam_tx0_send}
\varphi_{tx, 0}^{CAM} 
=&\left[1+P_S+P_S \sum_{z=1}^{T_C-2} \sum_{j=i+1}^{l-1} \frac{\left(1-P_S\right)^{\left(T_C-j\right)}}{\left[1-\left(1-P_S\right)^{\left(T_C-1\right)}\right]}\right.+
\\&\left.\sum_{j=0}^{T_C-1}\frac{\left(1-P_S\right)^{\left(T_C-j\right)}}{\left[1-\left(1-P_S\right)^{\left(T_C-1\right)}\right]}\right]^{-1}.
\end{aligned}
\end{equation}

Similarly, we can solve the steady-state probability of DENM message generator: 
\begin{equation}
\label{e_denm_idle_send}
\varphi_{Idle}^{DENM}=\frac{1}{\left(1+T_{avg}\right)\left(1-e^{-T/T_{avg}}\right)k} \varphi_{tx, 0}^{DENM}.
\end{equation}

When $i=tx$, each state, except for state $\varphi_{tx, T_D-1}^{DENM}$, is given by:
\begin{equation}
\begin{aligned}
\label{e_denm_tx_send}
\varphi_{tx, j}^{DENM} &=P_S\left[\frac{(k-1)}{k} \varphi_{tx, 0}^{DENM}+\sum_{z=j+1}^{T_{D}-1} \varphi_{t', z}^{DENM}\right] 
\\ & \quad for \ j \in\left[1, T_D-2\right].
\end{aligned}
\end{equation}
For $j = T_D-1$, the steady state probability is $\varphi_{tx, T_D-1}^{DENM}=\varphi_{tx, 0}^{DENM}P_S(k-1)/k$.
When $i=tx'$, the state is given by:
\begin{equation}
\begin{aligned}
\label{e_denm_tx'_send}
\varphi_{tx', j}^{DENM} & =  \varphi_{tx, 0}^{DENM}\frac{(k-1)}{k} \frac{\left(1-P_S\right)^{T_D-j}}{\left[1-\left(1-P_S\right)^{T_D-1}\right]}, 
\\ & \quad for \ j \in\left[0, T_D-1\right].
\end{aligned}
\end{equation}

By the property that the sum of the steady-state probabilities of DTMC is one, $\varphi_{tx, 0}^{DENM}$ can be obtained.

The steady-state probability of CAM and DENM generator is obtained, and the steady-state probability of device buffer queue and each state in queue can be obtained by joint probability solving. First, we solve the transition probability of two kinds of traffic respectively.

For CAM traffic, the transition probability of each state of the queue can be represented by:
\begin{equation}
\left\{\begin{array}{ll}
\label{e_cam_queue}
\alpha_{CAM}'=\varphi_{tx, 0}^{CAM}\left(1-P_S\right)+\varphi_{tx', 0}^{CAM} \hfill\smallskip\\
\alpha_{CAM}=\varphi_{tx', 0}^{CAM} \hfill \smallskip\\
\beta_{CAM}=\sum_{j=1}^{T_C-1} \varphi_{tx', j}^{CAM} P_S \hfill \smallskip
\end{array}\right.
\end{equation}

Similarly, we can obtain each transfer probability in the DENM traffic queue.

The queue model is one-dimensional DTMC. Using the detailed equilibrium equation, the steady-state solution of each state in the queue can be obtained:
\begin{equation}
\label{e_queue0}
\varphi_0=\left[1+\alpha'\left(\frac{1-\beta^{-m} \alpha^m}{\beta-\alpha}\right)\right]^{-1}
\end{equation}

\begin{equation}
\label{e_queuei}
\varphi_i=\frac{\alpha' \alpha^{i-1}}{\beta^i} \varphi_0, \quad for \  i \in(1, m)
\end{equation}

At the same time, the probability that the queue is empty is $P_{qe}=\varphi_0$.
The probability that a message is triggered at least once when the system is empty 
$P_{arr}=\varphi_{tx, 0}^{CAM}+\left(1-e^{-1/(T-T_{avg}/2)}\right)-\varphi_{tx, 0}^{CAM}\left(1-e^{-1/(T-T_{avg}/2)}\right)$.

\subsection{Model for NR-V2X Mode 2 with Re-Evaluation Mechanism}

The re-evaluation mechanism model is solved in parallel with the queue model. The queue model provides the probability of empty queue $P_{qe}$ and the probability of new message generation $P_{arr}$, and the re-evaluation mechanism model provides the probability of successful message transmission $P_S$. At the same time, according to \cite{WNBA2022DTMCV2X}, the probability that the system can successfully select candidate single-frame resources (CSR) is obtained from the number of vehicles and channel occupancy in the environment, which is represented by $P_{csr}$.

When solving the steady-state probability, the existence of re-evaluation mechanism leads to cross transitions between states. Therefore, virtual state $\varphi_S$ is set in the model for iterative solution, but it does not participate in steady-state probability.

For state $\varphi_{w,j}$ in the selection window, the relationship between states can be expressed as:
\begin{equation}
\label{e_sw-relation}
\left\{\begin{array}{l}
\varphi_{w, j} =\varphi_S \frac{1}{T_2-T_1}+\sum_{i=j}^{T_2-T_1-1} \varphi_{w, i}\left(1-P_{re}\right)^{i-1} \\
 \qquad \vdots \\
\varphi_{w, T_2-T_1-1} =\varphi_S \frac{1}{T_2-T_1}+P_{qne} P_{RC} \varphi_{1,0}
\end{array}\right.
\end{equation}

For the virtual state $\varphi_S$ in $(w,i)$, it can be expressed as:
\begin{equation}
\label{e_virtual_S}
\varphi_S =P_{re}\sum_{j=i+1}^{T_2-T_1-1}\frac{T_2-T_1}{T_j-T_3} \varphi_{w, j}+a \varphi_{Idle}+b\varphi_{1,0}, 
\end{equation}
where $a = (P_{qne}+P_{arr}-P_{qne}P_{arr})P_{csr}, b=P_{qe}\left(1-P_{RC}\right) P_{csr}$.

Therefore, we can get the steady-state probability of each state in the selection window as follows:
\begin{equation}
\label{e_re-evaluation-sw}
\varphi_{i, j}= \begin{cases}
   \varphi_{S} \frac{1}{T_2-T_1}+\sum_{i=j+1}^{T_2-T_1-1} \varphi_{w, i}\left(1-P_{re}\right)^{i-1}\\ 
   \qquad \vdots \\
   \varphi_{S} \frac{1}{T_2-T_1}+P_{qne} P_{qe} \varphi_{1,0}
                \end{cases} 
\end{equation}

After obtaining the above probability, the idle steady-state probability of the re-evaluation mechanism model can be obtained using the detailed equilibrium equation:
\begin{equation}
\label{e_re-evaluation-idle}
\varphi_{Idle} = \frac{{\left[{\left({1-{P_{RC}}}\right)\left({\frac{1}{{{P_{csr}}}}-1}\right)} \right]}}{{\left[{{P_{arr}} + {P_{qne}}\left({1-{P_{arr}}} \right)} \right]}}{\varphi _{w,0}}
\end{equation}

Then the system enters the semi-persistent scheduler of the message, and the effect of the reselect counter value on the steady-state probability of the state needs to be considered. At this time, the steady-state probability of each state is mainly composed of random selection probability, scheduling failure probability and the probability of transferring to this point after successful scheduling at the previous time, which can be expressed as:

\begin{equation}
\label{e_re-evaluation-sps}
\varphi_{i, j}= \begin{cases}
    \frac{\varphi_{w, 0}}{P_{qne}\left(1+R_U-R_L\right)} & for \ i=R_U, j \in\left[1, T_2-1\right] \hfill \smallskip
\\ \frac{\varphi_{w, 0}\left(R_U-i+1\right)}{P_{qne}^2\left(1+R_U-R_L\right)} & for \ i \in\left[R_L, R_U\right), j \in\left[1, T_2-1\right] \hfill \smallskip
\\ \frac{\varphi_{w, 0}\left(R_U-i+1\right)}{P_{qne}\left(1+R_U-R_L\right)} & for \ i \in\left[R_L, R_U\right], j=0 \hfill \smallskip
\\ \frac{\varphi_{w, 0}}{P_{qne}} & for \ i \in\left[1, R_L\right), j \in\left[0, T_2-1\right] \hfill \smallskip
                \end{cases} 
\end{equation}

\section{Simulation Results}

In this section, we describe the simulation settings (shown in Table.\ref{table_Simulation_Settings}) and metrics and evaluate the impact of the re-evaluation mechanism of MAC layer performance by comparing SPS algorithm with or without re-evaluation mechanism, using proposed model (described in Section \ref{section_SystemModel}), and baseline model (inspired by the modeling approach proposed in \cite{Cao2022Baseline}), respectively.


\begin{table}[!htbp]
		\caption{Simulation Settings For Re-Evaluation Mechanism} 
		\label{table_Simulation_Settings}
		\centering
		\begin{tabular}{rcc} 
			\toprule 
			Scenario & High intensity  &  Low intensity \\
			\midrule 
			CAM inter-arrival time $T_{C} $ & 30 ms & 100 ms \\
			\midrule 
            DENM fixed inter-time $T_{avg} $ & 10 ms& 50  ms\\
            \midrule 
            DENM inter-arrival time $T_{D} $ & 20 ms& 100  ms\\
            \midrule 
            Resource Reservation Interval $ RRI $ & 10 ms & 50 ms \\
            \midrule 
			Bandwidth & \multicolumn{2}{c}{20 MHz} \\
            \midrule 
			Subcarrier spacing $(SCS)$ & \multicolumn{2}{c}{30 kHz} \\
            \midrule 
			Number of Sub-channels $N_{subCH}$ & \multicolumn{2}{c}{4} \\
            \midrule 
			Re-select probability $P_{rk}$ & \multicolumn{2}{c}{0.4}\\
            \midrule 
			Blind retransmission frequency $k$ & \multicolumn{2}{c}{3} \\
            \midrule 
			Vehicle density $\rho$ & \multicolumn{2}{c}{ 50,100,150 vehicles/km} \\
			\bottomrule 
		\end{tabular}
	\end{table}

We have set the vehicle can utilize up to 4 sub-channels (of 12 RBs each) in each slot to transmit the generated messages with the MCS index 13. With this MCS, A CSR requires at least 4 RBs to transmit packets of 200 bytes. The parameters $T_1$, $T_2$ and $T_3$ configure the limits of the selection window. They are set equal to 2 slots, PDB, 5 slots, respectively.

\subsection{Baseline Model}
The Baseline Model\cite{Cao2022Baseline} is developed to evaluate NR-V2X Mode 2 when there is no re-evaluation mechanism. We apply this modeling approach to the setting  considered in this paper to compare the impact of the re-evaluation mechanism on NR-V2X scheduling. We then replace the notation $N_{sen}$ and $p$ in \cite{Cao2022Baseline} with the $N$ and $P_{rk}$, which stands for the total number of vehicle and the persistence probability, respectively. With this notation, the collision probability $P_{col}$ is written as:
\begin{equation}
\label{e_pcol_N_baseline}
P_{col}=\frac{1}{1+P_{rk}}\left[1-\left(1-\frac{1-P_{rk}}{\overline{RC}\cdot CSR_t}\right)^{N-1}\right],
\end{equation}
where $CSR_t$ is the number of CSRs available in the selection window and $\overline{RC}$ stands for the mean value of the RC.

We then replace the notation $t_s$ in \cite{Cao2022Baseline} with the $T_1$. With this notation, the average latency $E[T]$ is written as:
\begin{equation}
\label{e_latency_baseline}
E\left[T\right]=\frac{3RRI-T_{1}}{2} +\sum_{j=1}^{\infty} jRRIP_{col}(1-P_{col})^{j}.
\end{equation}

\subsection{Simulation Model}
The calculation of the collision probability $P_{col}$ is similar to \cite{WNBA2022DTMCV2X}. First, we compute the steady-state probability of state (1,0). The cycle time of the system at state (1,0) is $1/\varphi_{1,0}$. In the system resource selection window, the probability of the surrounding interference vehicle entering the state (1,0) is $1-\prod_{i=1}^{T_2} (1-1/(1/\varphi_{1,0}-i)))=p_c$. Then the probability that the launching vehicle will reselect resources and select the same resources at this time, that is, the collision probability is $p_c(1-P_{RC})/(CSR_t-N)$. Therefore, the relationship between the collision probability and the number of vehicles in the resource department can be obtained as follows:
\begin{equation}
\label{e_pcol_N_simulation}
P_{col} = 1-\left[ 1-\left [1-\prod_{i=1}^{T_2} (1-\frac{\varphi_{1,0}}{1-\varphi_{1,0}i})\right]   \frac{(1-P_{RC})}{(CSR_t-N)}\right]^{N}.
\end{equation}

The device level queue actually constitutes a one-dimensional Markov chain, thus the latency can be expressed as:
\begin{equation}
\label{e_latency_simulation}
E\left[T\right]=\sum_{i=1}^{m} i\varphi_{i}/\eta \alpha,
\end{equation}
where $\eta$ represents the ratio of the number of RB available to a vehicle in a time slot to the number of RB required by a message.

\subsection{Result Analysis}
\begin{itemize}
  \item Average Latency
\end{itemize}

\begin{figure}[thpb]
	\centering
    \vspace{-0.25cm}
	\includegraphics[scale=0.55]{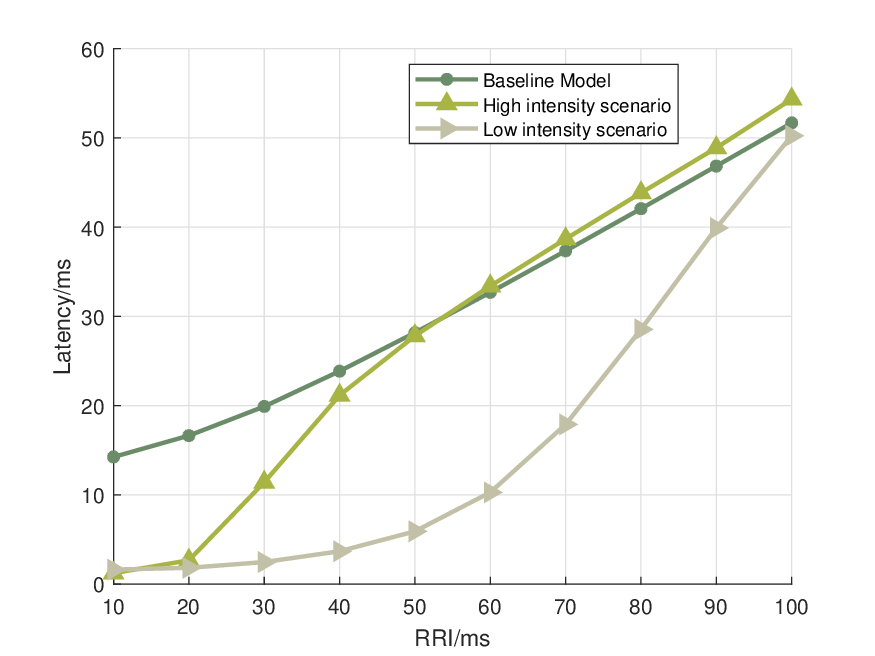}
    \vspace{-0.25cm}
	\caption{The performance of average latency versus RRI among different scenarios.}
	\label{fig_latency_re_base}
\end{figure}
Fig.\ref{fig_latency_re_base} depicts the relationship between average latency and RRI across various scenarios. In this figure, the baseline model represents the theoretical analysis of average latency in high-intensity scenarios where the re-evaluation mechanism is not activated. It can be observed that when the RRI is less than 50 ms, the system can achieve a lower latency. This is because the re-evaluation mechanism effectively decreases the collision probability and reduces the likelihood of re-transmission. 

As the RRI increases, the RBs that the system can select are more uniformly distributed within the selection window, leading to a slight increase in latency. In low-intensity scenario, this phenomenon is more pronounced, and the average delay escalates rapidly with an increase in the RRI.

%

\begin{itemize}
  \item Reliability and computational cost
\end{itemize}

\begin{figure}[thpb]
	\centering
    \vspace{-0.25cm}
	\includegraphics[scale=0.55]{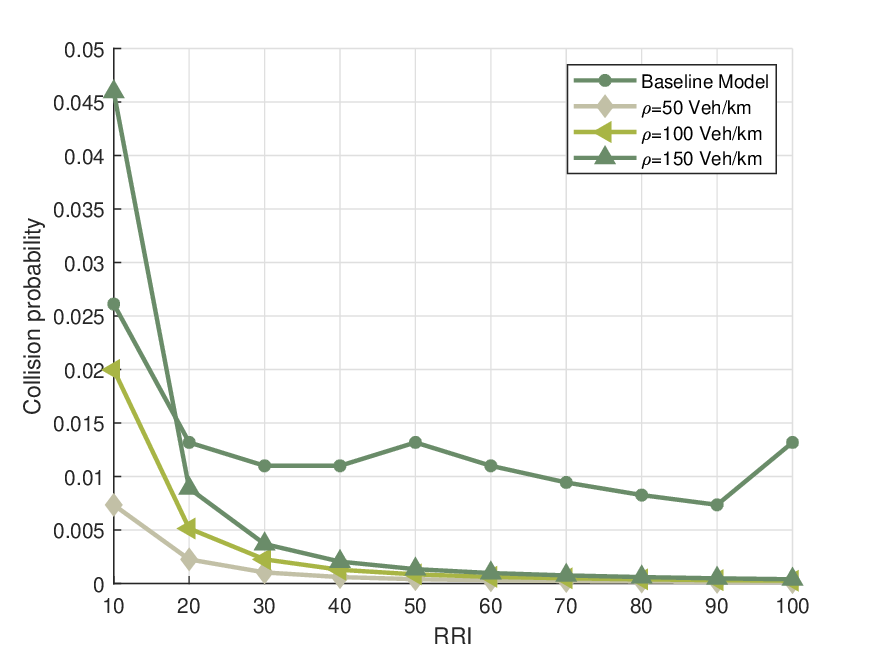}
    \vspace{-0.25cm}
	\caption{The performance of collision probability versus RRI among different vehicle density.}
	\label{fig_collision_probability_RRI_base}
\end{figure}

Fig.\ref{fig_collision_probability_RRI_base} depicts the relationship between the collision probability and RRI for different numbers of vehicles. As the RRI increases, the collision probability shows a tendency to decrease. This is mainly because as the RRI increases, the number of available RBs for the system rises, and the probability of different vehicles selecting the same RB decreases. When the RRI exceeds 50 ms, the difference in resource collision probability among different vehicle densities becomes less pronounced. 

The baseline model presented in the figure represents the theoretical analysis results at a density of 150 Veh/km without the re-evaluation mechanism being activated. In this baseline model, when the RRI equals 50 ms and 100 ms, the collision probability fluctuates as a result of the change in the value range of RC. Conversely, the implementation of the re-evaluation mechanism alleviates this volatility, reducing the collision probability by nearly an order of magnitude.

\begin{figure}[thpb]
	\centering
    \vspace{-0.25cm}
	\includegraphics[scale=0.55]{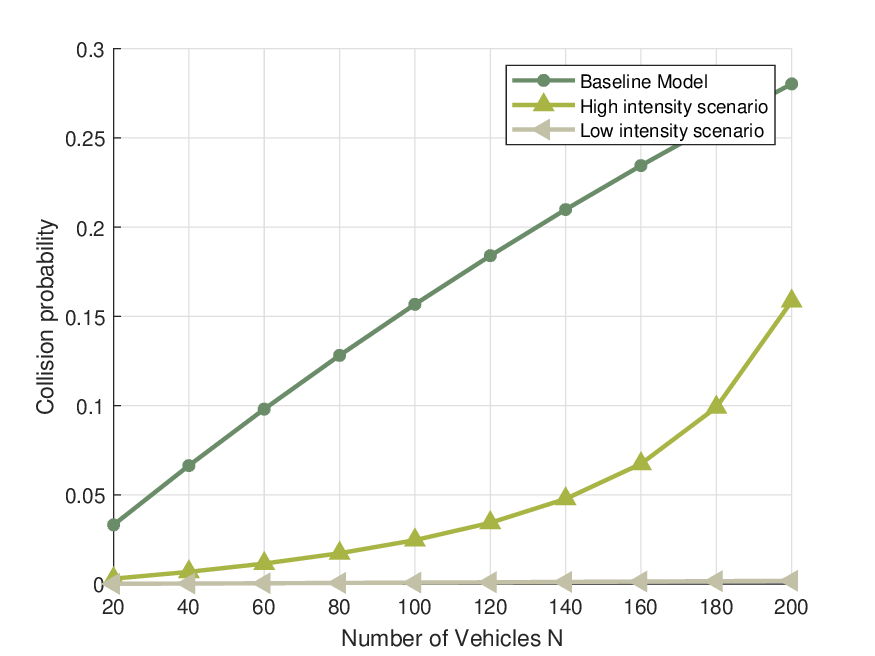}
    \vspace{-0.25cm}
	\caption{The performance of collision probability versus vehicles number N among different scenarios.}
	\label{fig_collision_probability_re_base}
\end{figure}

Fig. \ref{fig_collision_probability_re_base} depicts the relationship between the collision probability and the number of vehicles. The baseline model presented in the figure represents the outcomes of the theoretical analysis in a high-intensity scenario in which the re-evaluation mechanism is not activated. Through comparison, it can be observed that the re-evaluation mechanism can significantly decrease the collision probability. Nevertheless, in high-intensity scenarios when the volume of aperiodic traffic increases, the effectiveness of the re-evaluation mechanism is less promising, and it becomes difficult to meet the requirements of the ITS. In the future, a diverse range of methods should be contemplated to further lower the collision probability.


\section{Conclusion}
In this paper, we have proposed an analytical model of NR-V2X Mode 2 with re-evaluation mechanism.
We first construct new variants of message - generation models for CAM and DENM, along with a device - level queue model, to simulate variable V2X traffic scenarios. Subsequently, the established NR-V2X Mode 2 model is utilized to schedule messages. The entire system is composed of four inter-related DTMCs, each of which is an aperiodic, irreducible and recurrent chain. By solving the overall steady-state probability of the system, we assess the average latency, collision probability, and other indices.
Finally, numerical results demonstrate that the re-evaluation mechanism significantly reduces the probability of resource collision (less than 99\%). However, this reduction is achieved at the cost of increased average latency. Additionally, under the requirement of mandating the use of re-evaluation mechanisms, it remains worthwhile to consider reliability and latency when adjusting the timing of their application.



\bibliographystyle{IEEEtran}
\bibliography{ref}

\begin{thebibliography}{10}
\providecommand{\url}[1]{#1}
\csname url@samestyle\endcsname
\providecommand{\newblock}{\relax}
\providecommand{\bibinfo}[2]{#2}
\providecommand{\BIBentrySTDinterwordspacing}{\spaceskip=0pt\relax}
\providecommand{\BIBentryALTinterwordstretchfactor}{4}
\providecommand{\BIBentryALTinterwordspacing}{\spaceskip=\fontdimen2\font plus
\BIBentryALTinterwordstretchfactor\fontdimen3\font minus
  \fontdimen4\font\relax}
\providecommand{\BIBforeignlanguage}[2]{{%
\expandafter\ifx\csname l@#1\endcsname\relax
\typeout{** WARNING: IEEEtran.bst: No hyphenation pattern has been}%
\typeout{** loaded for the language `#1'. Using the pattern for}%
\typeout{** the default language instead.}%
\else
\language=\csname l@#1\endcsname
\fi
#2}}
\providecommand{\BIBdecl}{\relax}
\BIBdecl

\bibitem{Shin2023LTEtoNR}
C.~Shin, E.~Farag, H.~Ryu, M.~Zhou, and Y.~Kim, ``Vehicle-to-everything (v2x)
  evolution from 4g to 5g in 3gpp: Focusing on resource allocation aspects,''
  \emph{IEEE Access}, vol.~11, pp. 18\,689--18\,703, 2023.

\bibitem{Lu2024LinkNR}
L.~Lusvarghi, B.~Coll-Perales, J.~Gozalvez, and M.~L. Merani, ``Link level
  analysis of nr v2x sidelink communications,'' \emph{IEEE Internet of Things
  Journal}, vol.~11, no.~17, pp. 28\,385--28\,397, 2024.

\bibitem{Bankov2023NRtraffic}
D.~Bankov, E.~Khorov, A.~Krasilov, and A.~Otmakhov, ``Analytical model of 5g
  v2x mode 2 for sporadic traffic,'' \emph{IEEE Wireless Communications
  Letters}, vol.~12, no.~8, pp. 1449--1453, 2023.

\bibitem{Molina2022Reevaluation}
A.~Molina-Galan, B.~Coll-Perales, and J.~Gozalvez, ``Re-evaluation strategies
  for 5g nr v2x communications,'' in \emph{2022 IEEE 96th Vehicular Technology
  Conference (VTC2022-Fall)}, 2022, pp. 1--5.

\bibitem{Wang2022fixed}
X.~Wang, R.~A. Berry, I.~Vukovic, and J.~Rao, ``A fixed-point model for
  semi-persistent scheduling of vehicular safety messages,'' in \emph{2018 IEEE
  88th Vehicular Technology Conference (VTC-Fall)}, 2018, pp. 1--5.

\bibitem{Bal2023E2E}
B.~Coll-Perales, M.~C. Lucas-Estañ, T.~Shimizu, J.~Gozalvez, T.~Higuchi,
  S.~Avedisov, O.~Altintas, and M.~Sepulcre, ``End-to-end v2x latency modeling
  and analysis in 5g networks,'' \emph{IEEE Transactions on Vehicular
  Technology}, vol.~72, no.~4, pp. 5094--5109, 2023.

\bibitem{WNBA2022DTMCV2X}
G.~P. Wijesiri N. B.~A, J.~Haapola, and T.~Samarasinghe, ``The effect of
  concurrent multi-priority data streams on the mac layer performance of ieee
  802.11p and c-v2x mode 4,'' \emph{IEEE Transactions on Communications},
  vol.~70, no.~1, pp. 592--605, 2022.

\bibitem{3gpp2019service}
3GPP, ``Service requirements for enhanced v2x scenarios,'' \emph{Technical
  Specification (TS) 22.186, 3rd Generation Partnership Project (3GPP)}, 2019.

\bibitem{Romeo2020kDENM}
F.~Romeo, C.~Campolo, A.~Molinaro, and A.~O. Berthet, ``Denm repetitions to
  enhance reliability of the autonomous mode in nr v2x sidelink,'' in
  \emph{2020 IEEE 91st Vehicular Technology Conference (VTC2020-Spring)}, 2020,
  pp. 1--5.

\bibitem{3gpp2022MAC}
3GPP, ``Medium access control (mac) protocol specification,'' \emph{Technical
  Specification (TS) 38.321, 3rd Generation Partnership Project (3GPP)}, 2022.

\bibitem{3gpp2021data}
------, ``Physical layer procedures for data,'' \emph{Technical Specification
  (TS) 38.214, 3rd Generation Partnership Project (3GPP)}, 2021.

\bibitem{Cao2022Baseline}
L.~Cao, H.~Yin, R.~Wei, and L.~Zhang, ``Optimize semi-persistent scheduling in
  nr-v2x: An age-of-information perspective,'' in \emph{2022 IEEE Wireless
  Communications and Networking Conference (WCNC)}, 2022, pp. 2053--2058.

\end{thebibliography}


\end{document}